\documentclass[preprint,aps,tightenlines,showkeys,nofootinbib,superscriptaddress]{revtex4-2}
\usepackage{color}
\usepackage{enumerate}
\usepackage{latexsym}
\usepackage{graphicx}
\def\be{\begin{equation}}
\def\ee{\end{equation}}
\def\bea{\begin{eqnarray}}
\def\eea{\end{eqnarray}}

\begin{document}

%\preprint{arXiv:2202.xxxxx [hep-th]}
\today\\

\title{Deflection of light by magnetars in the generalized Born-infeld electrodynamics}

\author{Jin Young Kim \footnote{E-mail address: jykim@kunsan.ac.kr} }
\affiliation{Department of Physics, Kunsan National University,
Kunsan 54150, Korea}

\begin{abstract}
We study the deflection of light by a magnetic dipole field in the generalized Born-Infeld electrodynamics. 
Using the effective index of refraction and the trajectory equation based on geometric optics, we compute the weak bending angle of light passing on the equator of the magnetic dipole. In the limit where the classical Born-Infeld parameter is infinite, the bending angle agree with the one computed from the Euler-Heisenberg Lagrangian. We also compute the bending angle using the geodesic equation of the effective metric induced by a massive object with magnetic dipole. In the massless limit the bending angle agrees with the computation using the trajectory equation. We apply the result to magnetars to estimate the order of magnitude for the bending angle. 

\end{abstract}

%\pacs{42.25.Dd, 12.20.-m, 12.90.+b }

\keywords{deflection of light, Born-Infeld electrodynamics, magnetar}

\maketitle

\newpage

\section{Introduction}

The classical Born-Infeld action was introduced to solve the infinite self-energy of a point charge in Maxwell's electrodynamics \cite{Born, BornInfeld}. 
Regardless of the existence of classical Born-Infeld electrodynamics, in the strong field regime, the nonlinear electrodynamic effects emerge from the vacuum polarization of quantum electrodynamics. The one-loop effective action is known as the Euler-Heisenberg Lagrangian \cite{HeisenbergEuler,Schwinger}. In the nonlinear electromagnetic theory, the speed of light in the presence of strong electric or magnetic field background is not constant and depends on the field strengths. In the classical Born-Infeld theory, the electric  field strength that the nonlinear effect becomes significant is estimated of the order $10^{20} {\rm V/m}$ \cite{Jackson} corresponding to magnetic induction of the order $10^{11} {\rm T}$. In quantum electrodynamics, the relevant field strength is $B_c = m_e^2 c^2 / e \hbar \simeq 4.4 \times 10^{9} {\rm T}$ \cite{Schwinger}. 

There have been many ground laboratory experiments to observe the quantum electrodynamic nonlinearity \cite{BRST,PVLAS,BMV}. 
However, at strong field strengths unaccessible on Earth, it seems that charged or magnetized astronomical objects might give a deeper understanding of nonlinear electrodynamics. 
Various nonlinear electrodynamic models are adopted to many topics of astrophysics and cosmology \cite{rasheed, garcia, breton, camara}. 
Interests in astrophysical compact objects with extreme electromagnetic field are increasing, especially, in charged black holes and  magnetars. 

In geometric optics a light ray passing regions with nonuniform index of refraction bends towards higher index of refraction. The light ray passing the massive astronomical objects also bends by the general relativistic effect. The path of light is described by the geodesic equation. The light bending by the gradient in the gravitational field of a massive object known as the gravitational lensing is a very useful tool in astronomy and astrophysics. When the light ray is passing not close to the compact objects, the deflection is small and one can compute the bending angle in weak field approximation \cite{Weinberg}.
If the compact objects have strong electric or magnetic field, the path of light is also affected by the nonlinear interaction between the light and the strong background field. One can compute the trajectory from the geodesic equation including both mass and charge (or magnetic dipole). If one focuses on the bending by purely nonlinear electromagnetic effect, one can compute the bending angle in the geometric optics formalism \cite{dds,KimLee1,KimLee2}. 

In the previous work of the author, the deflection angle of light by Einstein-Born-Infeld black hole was computed in the weak bending limit \cite{kim2021}. The key point is that, due to nonlinear electrodynamic effects, photons do not propagate along the null geodesics of the background spacetime. They propagate along the so-called effective geodesics \cite{Plebanski, Novello, Breton}. Magnetic fields are more useful than electric fields in many areas of high energy astrophysics, for example, magnetized neutron stars. The purpose of this paper to compute the bending angles of light induced by strong magnetic field of magnetars in the generalized Born-Infeld electrodynamics. 

The organization of the paper is as follows. In Sec. II, we begin by briefly reviewing the generalized Born-Infeld action and the equations of motion. We consider the propagation of light in the magnetic field background induced by a magnetic dipole. Using the trajectory equation based on geometric optics, we compute the weak bending angle when the light is passing on the equatorial plane of the magnetic dipole. In the limit where the classical Born-Infeld parameter is infinite we compare the bending angle with the result obtained from the Euler-Heisenberg Lagrangian. In Sec. III, we consider a massive object possessing a magnetic dipole moment to see the general relativistic effects. From the static axially symmetric solution for the metric and the electromagnetic four-potential, we compute the bending angle using the effctive geodesics on the equatorial plane. In the massless limit, we show that the result agrees with the bending angle computed from the trajectory equation. In Sec. IV, as a possible application of the result to astrophysics, we estimate the 
order of magnitude of the bending angle for magnetars. Finally in Sec. V, we summarize and discuss our results. 

\section{Bending angle using geometric optics} 

%\subsection{Generalized Born-Infeld action and index of refraction} 

The classical Born-Infeld action is described by the Lagrangian
\be
 {\cal L}_{BI} = \beta_0^2 \left ( 1- \sqrt{ 1 + \frac{2 S}{\beta_0^2} - \frac{P^2}{\beta_0^4} } \right ), 
 \label{cbilagran}
\ee
where $\beta_0$ is the classical Born-Infeld parameter characterizing the possible maximum value of the field strength, $S$ and $P$ are Lorentz-invariants defined by
\be
 S = \frac{1}{4} F_{\mu \nu} F^{\mu \nu} = \frac{1}{2} ( {\bf B}^2 -  {\bf E}^2 )  , 
 ~~~ P = \frac{1}{4} F_{\mu \nu} {\tilde F}^{\mu \nu} =  {\bf E} \cdot {\bf B} ,
 \label{defSP}
\ee
$F_{\mu \nu} = \partial_\mu A_\nu - \partial_\nu A_\mu $ is the field strength tensor,  
and ${\tilde F}_{\mu \nu}  = \frac{1}{2} \epsilon_{\mu \nu \alpha \beta} F^{\alpha \beta} $ is its dual tensor. 
In the limit $\beta_0 \to \infty$ the above action reduces to the Maxwell action. 
Here we use the unit system with $1 / 4 \pi \epsilon_0 = \mu_0 / 4 \pi = \hbar = c = 1$. 

In quantum electrodynamics the one-loop correction of the vacuum polarization also induces non-linear terms,  the Euler-Heisenberg Lagrangian,
\be 
 {\cal L}_{EH} = \frac{8}{45}  \frac{\alpha^2}{m_e^4} S^2 + \frac{14}{45}  \frac{\alpha^2}{m_e^4} P^2.  
 \label{ehlag}
\ee 
As mentioned before the classical Born-Infeld parameter $\beta_0$ is estimated much larger than $E_c$.  Because the classical Born-Infeld Lagrangian is the same as the Maxwell Lagrangian for large $\beta_0$, the quantum correction of the classical Born-Infeld electrodynamics to the leading order would be the same as the Euler-Heisenberg Lagrangian. Thus, the effective action of the generalized (classical + quantum) Born-Infeld electrodynamics can be written as \cite{Kruglov10}
\be
 {\cal L} = \beta^2 \left ( 1- \sqrt{ 1 + \frac{2 S}{\beta^2} - \frac{P^2}{\beta^2 \gamma^2} } \right ),
 \label{GBIaction}
\ee
where $\beta$ and $\gamma$ are defined as
\be 
\frac{1}{ \beta^2} \equiv \frac{1}{ \beta_0^2} +  \frac{16}{45}  \frac{\alpha^2}{m_e^4} , ~~~
       \frac{1}{ \gamma^2} \equiv \frac{1}{ \beta_0^2} + \frac{28}{45}  \frac{\alpha^2}{m_e^4}.  
 \label{betagamma}
\ee      
The equations of motion, obtained from the Euler-Lagrange equation and the Bianchi identity, are   
\be 
\partial_\mu \left [ \frac{1} {\cal R}  \left ( F^{\mu \nu} - \frac{P}{\gamma^2} {\tilde F}^{\mu\nu} \right ) \right ] = 0 ,
\label{eqsofmotion}
\ee
\be 
\partial_\mu  {\tilde F}^{\mu\nu}  = 0 ,
\label{bianchieq}
\ee
where 
\be
 {\cal R} = \sqrt{ 1 + \frac{2 S}{\beta^2} - \frac{P^2}{\beta^2 \gamma^2} } .
 \label{defcalR}
\ee

The propagation of light in a static uniform magnetic field, assuming that the background magnetic field ($\bar {\bf B}$) is much stronger than the photon's field and is perpendicular to the direction of photon, can be described by the effective index of refraction given by \cite{Kruglov10}
 \be 
n_{\bot} =  \left ( 1 + \frac{ {\bar {\bf B}}^2 }{\beta^2} \right)^{\frac{1}{2}}  , ~~~
n_{\parallel} =  \left ( 1 + \frac{ {\bar {\bf B}}^2 }{\gamma^2} \right)^{\frac{1}{2}}. 
\label{effindrefmag} 
\ee
where $n_{\bot}$ ($n_{\parallel}$ ) is for the light mode polarized perpendicular (parallel) to the background magnetic field. Assuming that the photon is traveling in the $x$-direction and $\bar {\bf B}$ is in the $z$-direction, $n_{\bot} = n_y$ and $n_{\parallel}= n_z$. 
Note that in the classical Born-Infeld theory where $\beta = \gamma$ the two indices of refraction are the same. However, they are different including the quantum correction so that vacuum birefringence effect is relevant. 

%\subsection{Trajectory equation }

Because the effective index of refraction depends on the background field, the light ray can be bent continuously when the background field is non-uniform. We consider one of the simple cases to compute the bending angle within the geometric optics formalism. In the weak bending approximation where the impact parameter is large compared with the radius of the lensing object, the trajectory equation based on Snell's law can be written as 
\cite{KimLee1,KimLee2}
 \be
 \frac{d{\bf u}}{ds}=\frac{1}{n}({\bf u}\times {\nabla}n)\times {\bf u},  \label{trajeceq}
 \ee
where $s$ is the differential distance $ds=|d\vec{\bf r}|= \sqrt{dx^2 + dy^2  + dz^2 }$ and ${\bf u}= {d {\bf r}}/{ds}$ is the unit vector in the direction of light. 
It has been confirmed that the bending angle computed from this trajectory equation exactly agrees with the one obtained from the eikonal equation \cite{dds,KimLee2}. 

As far as the index of refraction is close to one, Eq. (\ref{trajeceq}) can be written as 
 \be
 \frac{d{\bf u}}{ds}=\frac{1}{n}({\bf u}_0 \times {\nabla}n)\times {\bf u}_0 ,  \label{trajeceq2}
 \ee
 where ${\bf u}_0$ is the direction of the incoming light. 
When the light ray is coming from $x=-\infty$ and moving to $x= + \infty$ with impact parameter $b$, ${\bf u}_0=(1,0,0)$,
the trajectory equation in the leading order can be written as
 \be
 \frac{d^2x}{ds^2} = 0, ~~~ \frac{d^2 y}{ds^2} =\frac{1}{n}  \frac{\partial n}{\partial y},
 ~~~ \frac{d^2 z}{ds^2} = \frac{1}{n}  \frac{\partial n}{\partial z}  .    \label{trajeceq3}  
 \ee
The first equation gives $ds=dx$ and we have
\be
 \frac{d^2 y}{dx^2} =\frac{1}{n}  \frac{\partial n}{\partial y},
 ~~~ \frac{d^2 z}{dx^2} = \frac{1}{n}  \frac{\partial n}{\partial z}  .    \label{trajeceq4}  
 \ee
We will use Eq. (\ref{trajeceq4}) to compute the bending angles of light for the effective indices of refraction given by 
Eq. (\ref{effindrefmag}). 
If we take $\varphi $ as the angle between $\bf u$ and the $x$ axis, the slope of the trajectory is $ y^\prime = dy /dx = \tan \varphi \simeq \varphi$. Then the bending angle can be obtained by integration 
 \be 
 \Delta \varphi =  y' (\infty) - y'(-\infty) .  \label{delvarphi}
\ee
 
%\subsection{ Bending angle by a magnetic dipole}

Now we consider the bending of light by a magnetic dipole.
The magnetic induction induced by a magnetic dipole $\bf m$, located at the origin, is 
\be
{\bar {\bf B}} =  \frac { 3 ({\bf m} \cdot {\bf r} ) {\bf r} }{r^5} - \frac{\bf m}{r^3 }  .
\label{dipolefield}
\ee
Contrary to the electric field by a Coulomb charge which is isotropic, the magnetic induction by a magnetic dipole is anisotropic. 
So the bending angle depends on the orientation of magnetic dipole relative to the incoming light ray. For simplicity of computation, we consider the simple case when the light ray is passing in the equatorial plane of the magnetic dipole (Fig. 1). 

\begin{figure}
\begin{center}
\includegraphics[height=8cm,keepaspectratio]{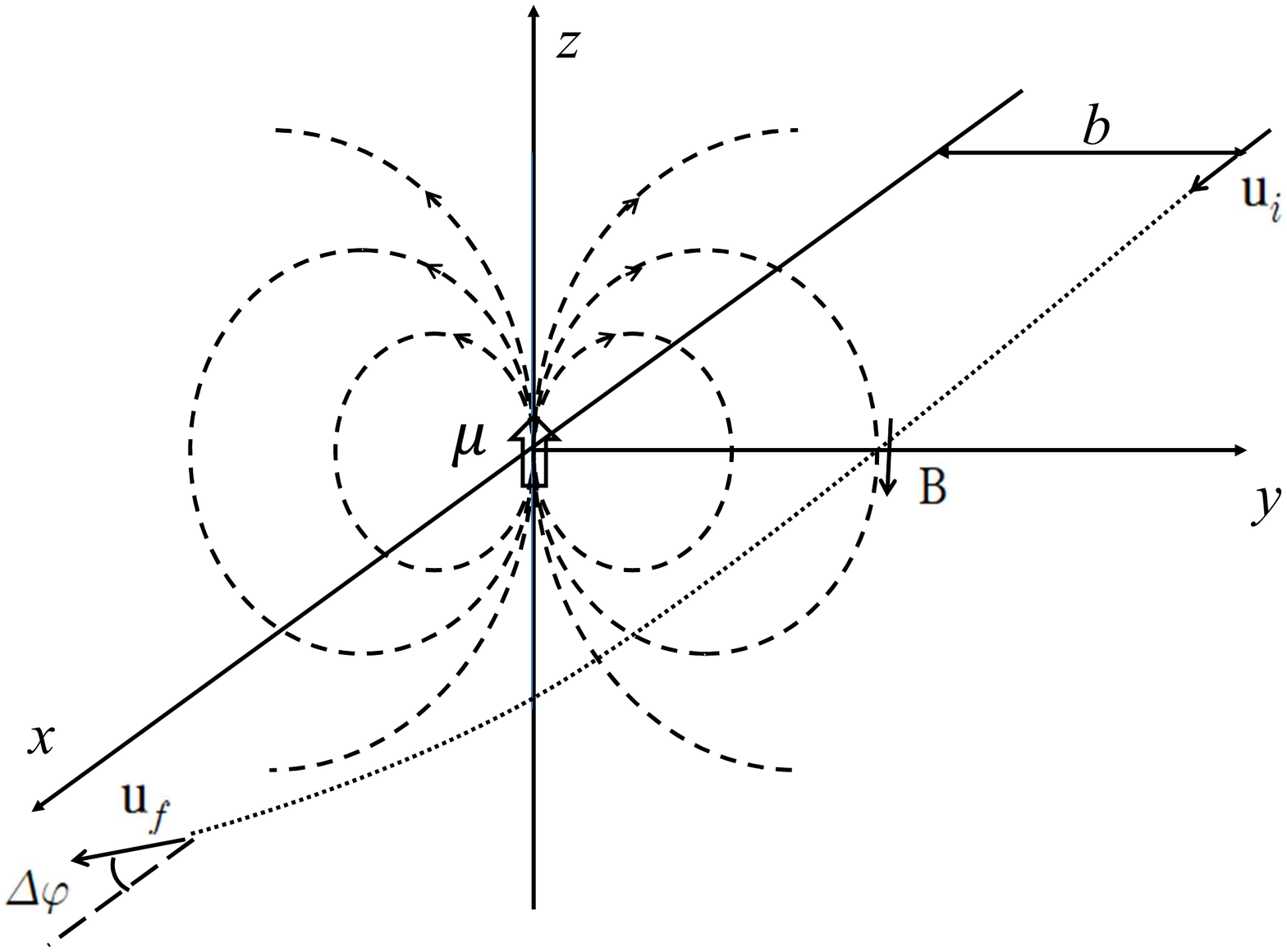}
\caption{Schematic of light bending when a light ray passes on the equatorial plane of a magnetic dipole } \label{fig1}
\end{center}
\end{figure}

Taking the direction of magnetic dipole as the $z$-axis ${\bf m} = \mu \hat z$, the magnetic induction on the equator is given by
\be
{\bar {\bf B}} =-  \frac {\mu}{ r^3} {\hat z},
\label{dipolefieldz}
\ee
where $r = \sqrt{x^2 +y^2}$. The indices of refraction can be written as 
\be 
n_{\bot} = \left ( 1 + \frac{ \mu^2 }{\beta^2 r^6} \right)^{\frac{1}{2}}  , ~~~
n_{\parallel} = \left ( 1 + \frac{ \mu^2 }{\gamma^2 r^6} \right)^{\frac{1}{2}}. 
\label{effindrefmagunit}
\ee
Because $\bar {\bf B}= {\bar {\bf B}}(x,y)$ on the equator, the index refraction does not depend on the coordinate $z$. Thus, there is no bending in the $z$-direction from Eq. (\ref{trajeceq4}). The bending angle in the $y$-direction for the perpendicular mode can be computed from
\be
y^{\prime \prime} = -3 \frac{ \mu^2}{\beta^2} \frac{y}{r^8} \frac{ 1} { 1 + \frac{\mu^2}{\beta^2} \frac{1}{r^6} }.  \label{trajb}
\ee
Because the slope $dy/dx = y'$ of the trajectory $y(x)$ is the direction of light, the boundary conditions that the incoming ray with impact parameter $b$ is in the direction of $+x$-direction can be written as 
 \be
 y(-\infty)=b, ~~~~ y'(-\infty)=0 . \label{inicond} 
 \ee
Integrating Eq. (\ref{trajb}) with the boundary conditions, the bending angle obtained from Eq. (\ref{delvarphi}) is 
\be
 \Delta \varphi_{\bot} = \int_{-\infty}^{\infty} y^{\prime \prime} dx 
 = -3 \pi \left[ 1 - \frac{1}{3} \frac{1}{\sqrt{ 1 +a^2}} -\frac{\sqrt{2}}{3} \frac{1}{ ( 1 -a^2 +a^4 )^{1/4}} 
\left ( 1 + \frac{1 - a^2 /2}{ \sqrt{1 -a^2 +a^4}} \right )^{1/2} \right ],    \label{bendangdipole}
\ee
where $a$ is defined as 
\be
a \equiv \left ( \frac{\mu}{\beta b^3} \right )^{1/3}.
\ee
The bending angle for the parallel mode can be obtained by replacing $\beta$ with $\gamma$ in Eq. (\ref{trajb}). 

Let us check whether Eq. (\ref{bendangdipole}) reduces to the bending angle computed from the Euler-Heisenberg Lagrangian when the classical Born-Infeld parameter becomes infinite. 
By series expansion for small $a$, we obtain 
\be
 \Delta \varphi_{\bot} = - \frac{15 \pi}{16} \frac{\mu^2}{\beta^2 b^6} + \cdots .  \label{leadintermgoc}
\ee
The negative sign means that the bending occurs toward the magnetic dipole. In the limit $\beta_0 \to \infty$, replacing $1/\beta^2$ with $16 \alpha^2 /45 m_e^4$ from Eq. (\ref{betagamma}), the leading order term becomes
\be
 \Delta \varphi_{\bot} = - \frac{ \pi}{3} \frac{\alpha^2}{m_e^4}\frac{\mu^2}{ b^6} .  \label{magdipben}
\ee
For the parallel mode we obtain $ \Delta \varphi_{\parallel} =  (7/4) \Delta \varphi_{\bot}$. 
This result exactly matches the bending angle obtained from the Euler-Heisenberg Lagrangian \cite{dds, KimLee2}. 

\section{Bending angle using geodesic equation}

In the previous section, we computed the bending angle by a magnetic dipole using the trajectory equation, which is based on geometric optics. If we are interested in the light bending by a compact astronomical object with strong magnetic field we have to consider the general relativistic effect. The bending of light by an electrically charged Born-Infeld black hole was computed using the geodesic equation \cite{kim2021}. In the limit where the mass of black hole is going to zero, it was confirmed that the bending angle agree with the one computed from the geometric optics formalism. Here we compute the bending angle of light by a compact object with mass and magnetic dipole moment. 

For simplicity and comparision with the result in the previous section, we consider the light bending in the equator of the static magnetic dipole. 
We start from the Einstein-Born-Infeld action
\be 
S = \int d^4 x \sqrt{-g} \left ( \frac{R}{16 \pi } + \frac{1}{4 \pi} {\cal L} \right ) , \label{EBIaction}
\ee
where $ {\cal L}$ is given by Eq. (\ref{GBIaction}). We use the unit system with $1 / 4 \pi \epsilon_0 = \mu_0 / 4 \pi = G = c = 1$. 
The equations of motion can be obtained by varying the action with respect to $g_{\mu \nu}$ and $A_\mu$ with the Bianchi identity. Since we consider only the background magnetic component of $F_{\mu \nu}$ is nonzero (${\bar {\bf  B}} \ne 0 ,~{\bar {\bf  E}} = 0$), there is no contribution from the invariant $P$. In this case the equations of motion are given by
\bea
 R_{\mu \nu} - \frac{1}{2} g_{\mu \nu} R &&= 8 \pi T_{\mu \nu} , \\
 \nabla \left ( \frac {F_{\mu \nu} }{ \sqrt { 1 + \frac{2S} {\beta^2 }} } \right )&&=  0, 
\eea
where $T_{\mu \nu}$ is the energy momentum tensor given by
\be 
T_{\mu \nu} = \frac{1}{4 \pi} \left [ \beta^2 \left ( 1 - \sqrt{ 1 + \frac{2S}{\beta^2} } \right ) g_{\mu \nu}
  + \frac{F_{\mu \alpha} F^\alpha_{~~\nu}}{ \sqrt{ 1 + \frac{2S}{\beta^2} } } \right ] . 
\ee

These field equations have axially symmetric solution for the magnetic dipole moment $ {\vec \mu} = \mu \hat z$ at the center. 
Because of the axial symmetry, it is convenient to use the cylindrical coordinates, $x^\mu = ( t, r, z, \phi)$, with the following ansatzs for the metric and the electromagnetic four-potential
\bea
ds^2 &=& g_{\mu \nu} dx^\mu dx^\nu =  f(r, z) dt^2 - g(r, z) ( d r^2 + dz^2 ) - r^2 h(r, z) d \phi^2, \label{gbkgr} \\
A_\mu &=& (0,0,0, -\psi) .     \label{A4vec} 
\eea
The power series solutions in powers of gravitational constant was obtained by Martin and Pritchett \cite{martinpritchett}.
Up to the first order in the gravitational constant, the solutions are given by  
\bea
 f(r, z) &=& 1 - \frac {2 GM} {X } + \frac{ G \mu^2 z^2 } {X^6 } , \\
g(r, z) &=& 1 + \frac {2 GM}{X } - \frac{ G \mu^2 (r^4 - 6 r^2 z^2 + 2 z^4 )} { 2X^8 } , \\
 h(r, z) &=& 1 + \frac {2 GM}{X } - \frac{ G \mu^2 z^2 } { X^6} , \\
\psi (r, z) &=&  \frac {\mu r^2 }{ X^3 } \left ( 1 +  \frac{ GM} {2 X }  \right ),
\eea
where $X$ is the spherical distance $X = \sqrt{r^2 + z^2}$, $M$ is the mass and we recover $G$ to show the order of powers of the gravitational constant. 
The quantum electrodynamic correction to first order for the magnetic potential, in a flat space-time background, has been computed in spherical polar coordinates \cite{Heyl}. Similarly the higher order corrections for the magnetic four-potential $\psi(r,z)$, in curved spacetime background, can also be considered in cylindrical coordinates.

In the linear Maxwell electrodynamics coupled to gravity, photons and gravitons follow the same null geodesic made by mass and magnetic dipole. However, in the nonlinear electrodynamics coupled to gravity, the null geodesic of the electromagnetic wave is different from the null geodesic of the gravitational wave due to the nonlinear coupling of the electromagnetic wave to the background electromagnetic field. 
The modification of light cone condition can be represented by \cite{Plebanski,Delorenci,Eiroa} 
\be
\left ( g^{\mu \nu} + 4 \frac{ {\cal L}_{SS} }{ {\cal L}_S } F^{\mu \alpha} F^\nu_{~~\alpha} \right ) k_\mu k_\nu = 0 ,
\ee
 where $k_\mu$ is the 4-vector of the propagating photon and ${\cal L}_S$ denotes the derivative of ${\cal L}$ with respect to $S$. 
The effective metric that makes $k_\mu$ a null vector is
\be
 {\tilde g}^{\mu \nu} = g^{\mu \nu} + 4 \frac{ {\cal L}_{SS} }{ {\cal L}_S } F^{\mu \alpha} F^\nu_{~~\alpha} . \label{geff}
\ee
Substituting Eqs. (\ref{gbkgr}) and (\ref{A4vec}) into Eq. (\ref{geff}), up to the quadratic order in Born-Infeld parameter, we obtain 
\be
{\tilde g}^{\mu \nu} = \pmatrix { g^{00}  & 0  & 0  & 0 \cr 
0 &  g^{11} ( 1 - \psi_r^2 /\beta^2 r^2 )  &  \psi_r \psi_z /\beta^2 r^2 & 0 \cr  
0&  \psi_r \psi_z /\beta^2 r^2&  g^{22}  ( 1 - \psi_z^2 /\beta^2 r^2 )  &0  \cr
0&  0& 0  & -g^{33} [  1 - (\psi_r^2 + \psi_z^2)  /\beta^2 r^2 ]  } ,
\ee
where $\psi_r = { \partial \psi} / \partial r$ and $\psi_z = { \partial \psi} / \partial z$. 

\begin{figure}
\begin{center}
\includegraphics[height=8cm,keepaspectratio]{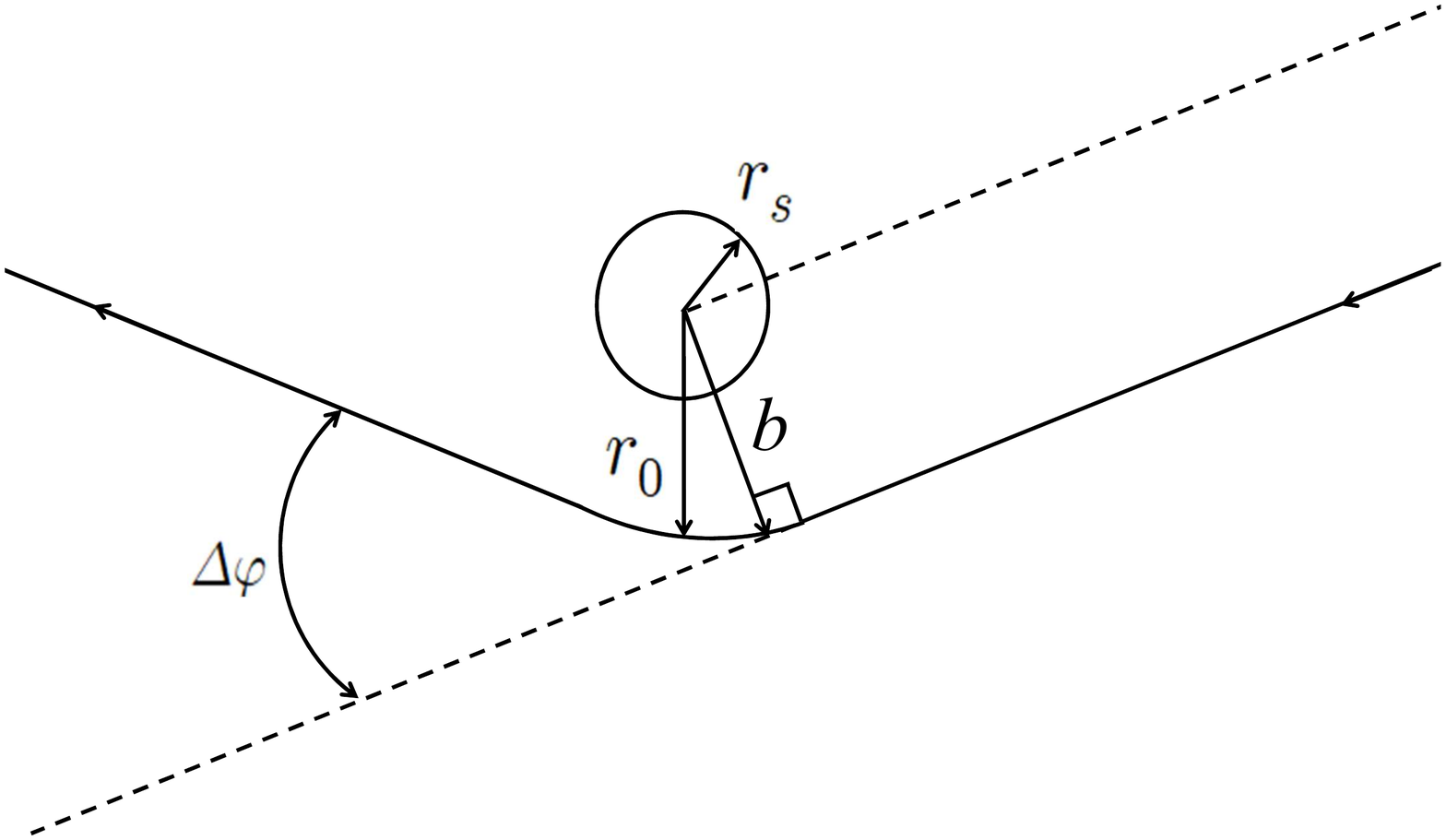}
\caption{Bending angle $\Delta \varphi$ of light ray on the equatorial plane of the magnetic dipole. The distance of the closest approach $r_0$ is the minimum of the radial coordinate $r$, $b$ is the impact parameter and $r_s$ is the radius of the compact object with magnetic dipole.} \label{fig2}
\end{center}
\end{figure}

The trajectory of weak bending on  the equatorial plane ($z = 0 $) can be computed from the following effective metric, up to the first order in $G$ and second order in $\mu$,
\be 
ds^2_{eq} = B(r) dt^2 - A(r) dr^2 - C(r)  d \phi^2 ,  \label{metriconequator}
\ee
where
\bea
 B(r) &=& 1 - \frac {2 GM} {r }  , \\
 A(r) &=& 1 + \frac {2 GM} {r } - \frac{ G \mu^2} { 2r^4 }  +  \frac {\mu^2} {\beta^2 r^6 }, \\
 C(r) &=& 1 + \frac {2 GM} {r }  +  \frac {\mu^2} {\beta^2 r^6 } . 
\eea
Following the notation and procedure in \cite{Weinberg}, the bending angle of light coming from infinity can be obtained from (Fig. 2)
\be 
\Delta \varphi = 2 | \varphi (r_0 ) - \varphi_\infty | - \pi, 
 \label{weinberg855}
\ee
whete $r_0$ is the distance of closest approach and 
\be
 \varphi (r_0 ) =  \varphi_\infty + \int_{r_0 }^\infty
    \left [ \frac {A(r)} {C(r)} \right ]^{1/2}    \left [ \frac {C(r) B(r_0 )} {C(r_0 ) B(r)} -1 \right ]^{-1/2} dr .
  \label{weinberg866}
 \ee                        

Substituting Eq. (\ref{metriconequator}) into Eq. (\ref{weinberg866}), we have
\be
 \varphi (r_0 ) - \varphi_\infty = 
 \int_{0 }^1 \frac{dx}{\sqrt{1 - x^2}} 
\left [ 1 + \frac{GM}{ r_0} \frac{1}{1+x} - \frac{G \mu^2}{4 r_0^4} x^4 + \frac{\mu^2}{2 \beta^2 r_0^6} (1 +x^2 +x^4) + \cdots \right ], 
  \label{angleintegral}
 \ee  
where $x = r_0 /r$.
The integral yields 
\be
 \varphi (r_0 ) - \varphi_\infty = 
\frac{ \pi}{2} + \frac{2GM}{ r_0} - \frac{3 \pi}{64}  \frac{G \mu^2}{  r_0^4} + \frac{ 15 \pi}{32}\frac{\mu^2}{\beta^2 r_0^6} +\cdots .
  \label{resultofint}
 \ee  
Inserting Eq. (\ref{resultofint}) in Eq. (\ref{weinberg855}), we have
\be 
\Delta \varphi = 
\frac{4GM}{ r_0} - \frac{3 \pi}{32}  \frac{G \mu^2}{ r_0^4} + \frac{ 15 \pi}{16}\frac{\mu^2}{ \beta^2 r_0^6}. \label{bendingang}
\ee
The first two terms having the gravitational constant come from the general relativistic effects while the third term reflects the nonlinear electrodynamic effects. 
The relative sign of the second term is different from the first and the third terms. This means that the contribution of the magnetic dipole to the total bending angle by nontrivial geodesic is repulsive while the contribution by nonlinear electrodynamic effect is attractive. For the mode polarized parallel to the magnetic field ($z$-mode), $\beta $ is replaced by $\gamma$ in Eq. (\ref{bendingang}).
In the weak deflection limit where $r_0 \simeq b$, the third term in Eq. (\ref{bendingang}) exactly matches the leading order computed from geometric optics Eq. (\ref{leadintermgoc}). 

\section{Order-of-magnitude estimation for magnetars}

The strongest magnetic field in the present observed universe is observed at highly magnetized neutron stars known as magnetars. The magnetic field on the surface of magnetars is estimated to reach up to the order $10^{11} \rm T$ \cite{duncan, thompson}. Let us apply the results to magnetars for the order-of-magnitude estimation. For this purpose we decompose Eq. (\ref{bendingang}) according to the order of the impact parameter $b$ as
\be
\Delta \varphi = \Delta \varphi_1 - \Delta \varphi_2 + \Delta \varphi_3 ,
\ee
 where 
\be
 \Delta \varphi_1 = \frac{4GM}{b},  ~~ \Delta \varphi_2 = \frac{3 \pi}{32}  \frac{G \mu^2}{ b^4} , 
~~  \Delta \varphi_3 =  \frac{ 15 \pi}{16}\frac{\mu^2}{ \beta^2 b^6}. \label{varphi123}
\ee
From Eq. (\ref{betagamma}), in the limit $\beta_0 \to \infty$, $\Delta \varphi_3$ can be written as
\be
 \Delta \varphi_{3} =  \frac{ \pi}{3} \frac{\alpha^2}{m_e^4}\frac{\mu^2}{ b^6} .  \label{varphi3eh}
\ee
Restoring all normalized units, $\Delta \varphi_{1}, \Delta \varphi_{2}$ and $\Delta \varphi_{3}$ can be written as
\be
 \Delta \varphi_1 =  \frac{4 G M}{c^2 b} ,
\ee
\be
 \Delta \varphi_2 =  \frac{3 \pi}{32} \frac{G}{c^4} \frac{\mu_0} {4 \pi}  \frac{\mu^2}{b^4}  ,
\ee
\be
 \Delta \varphi_3 =  \frac{ \pi}{3} \frac{\alpha^2}{m_e^4} \left ( \frac{\hbar^3 \epsilon_0}{c^3} \right ) 
\left ( \frac{\mu_0}{4 \pi} \right )^2
\frac{\mu^2}{ b^6} .
\ee
It is more useful to express $\Delta \varphi_2$ and $\Delta \varphi_3$ in terms of the magnetic induction on the surface of the magnetar and QED critical field strength as 
\be
 \Delta \varphi_2 =  \frac{3 \pi}{32} \frac{G}{c^4} \frac{4 \pi}{\mu_0} B_s^2 \frac{r_s^6}{b^4} ,
\ee
\be
 \Delta \varphi_3 =  \frac{\alpha}{12} \frac{B_s^2}{B_c^2} \frac{r_s^6}{b^6}  ,  \label{phi3EH}
\ee
where $r_s$ is the radius of the magnetar, $B_s =(\mu_0 / 4 \pi)( \mu /r_s^3)$,
 $B_c = m_e^2 c^2 / e \hbar = 4.4 \times 10^9 {\rm T}$, and $\alpha = e^2 / 4 \pi \epsilon_0 \hbar c$.  

For the order-of-magnitude estimation for magnetar, we consider the mass and radius of the typical neutron star, 
$M= 1.4 M_{\rm sun} = 2.8 \times 10^{30} {\rm kg} $, $r_s = 10 {\rm km}$, and the maximal surface magnetic field $B_s = 10^{11} \rm T$. For these values the Schwarzschild radius is about $4.2 {\rm km}$. The possible maximal bending angle is obtained from a light ray glancing ($b \simeq r_s$) the equator of the magnetar as 
\be 
 \Delta \varphi_1 = 8.30 \times 10^{-1} {\rm rad}, ~~
 \Delta \varphi_2 = 2.42 \times 10^{-8} {\rm rad}, ~~
 \Delta \varphi_3 = 3.14 \times 10^{-1} {\rm rad }. 
\ee
In table 1, we list the numerical estimations of  $\Delta \varphi_1 $, $\Delta \varphi_2$ and $\Delta \varphi_3 $ for certain values of impact parameter. 
Near the equator of the magnetar, the bending caused by nonlinear electrodynamic effect $\Delta \varphi_3$ is comparable to gravitational lensing by mass $\Delta \varphi_1$.  Because $ \Delta \varphi_1 \propto b^{-1}$ while $ \Delta \varphi_3 \propto b^{-6}$, we can neglect the lensing by nonlinear electrobynamic effect away from the magnetar. For example, at $ b = 5 r_s$, we have 
\be 
 \Delta \varphi_1 = 1.67 \times 10^{-1} {\rm rad}, ~~
 \Delta \varphi_2 = 3.87 \times 10^{-11} {\rm rad}, ~~
 \Delta \varphi_3 = 2.01 \times 10^{-5} {\rm rad }. 
\ee

To see the effect of the finiteness of the classical Born-Infeld parameter, we plot the bending angle $|\Delta \varphi_{\bot}|$ given by Eq. (\ref{bendangdipole}) and the bending angle $\Delta \varphi_3$ given by Eq. (\ref{phi3EH}), corresponding to Euler-Heisenberg limit, compared with the gravitational bending angle $\Delta \varphi_1$ in Fig. 3. 
We also plot $|\Delta \varphi_{\bot}|$ and the total bending angle $\Delta \varphi (\simeq \Delta \varphi_1 + \Delta \varphi_3)$ in Eq. (\ref{bendingang}) to see the dependence on mass in Fig. 4. 

\begin{table}
\begin{center}
\includegraphics[height=8cm,keepaspectratio]{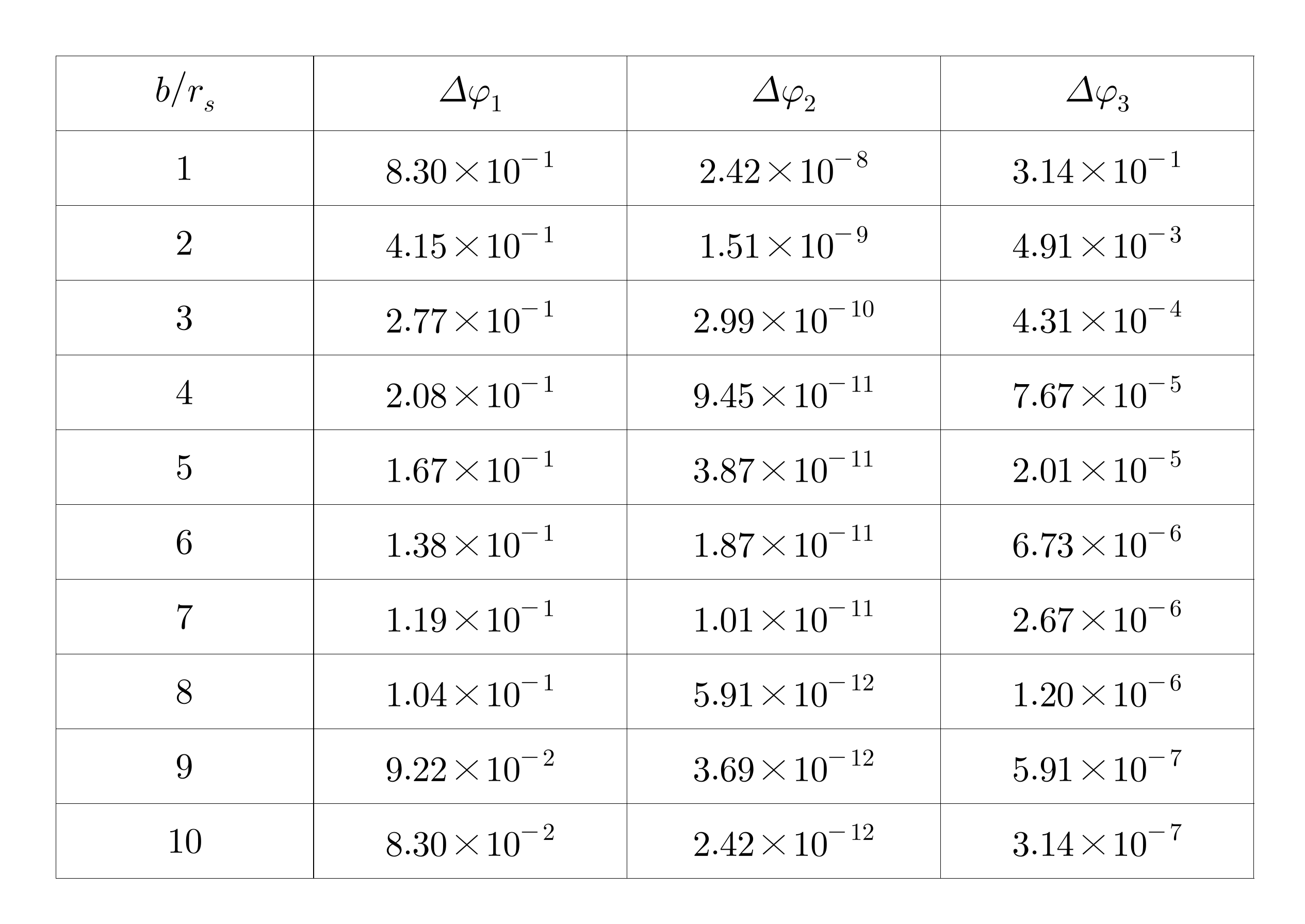}
\caption{Numerical estimation of each bending angles in Eq.  (\ref{varphi123}) for certain values of $r_s \le b \le 10 r_s $ with
$M= 1.4 M_{\rm sun}$, $r_s = 10 {\rm km}$ and $B_s = 10^{11} \rm T$. } \label{table1}
\end{center}
\end{table}

\begin{figure}
\begin{center}
\includegraphics[height=6cm,keepaspectratio]{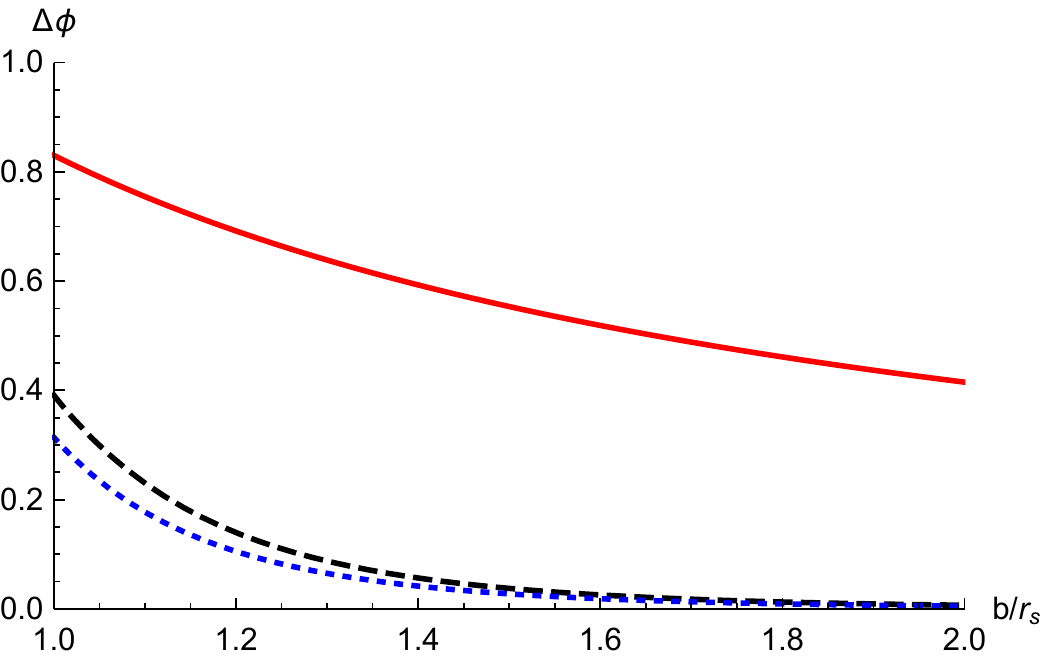}
\caption{The plots of $\Delta \varphi_1$ (solid), $|\Delta \varphi_{\bot}|$ (dashed), and $\Delta \varphi_3$ (dotted) for $\beta_0 = 5 \times 10^{11} \rm T$ with the same $M$, $r_s$ and $B_s$ as in Table 1. As $\beta_0$ becomes larger the dashed curve comes closer to the dotted curve. } \label{fig3}
\end{center}
\end{figure}

\begin{figure}
\begin{center}
\includegraphics[height=6cm,keepaspectratio]{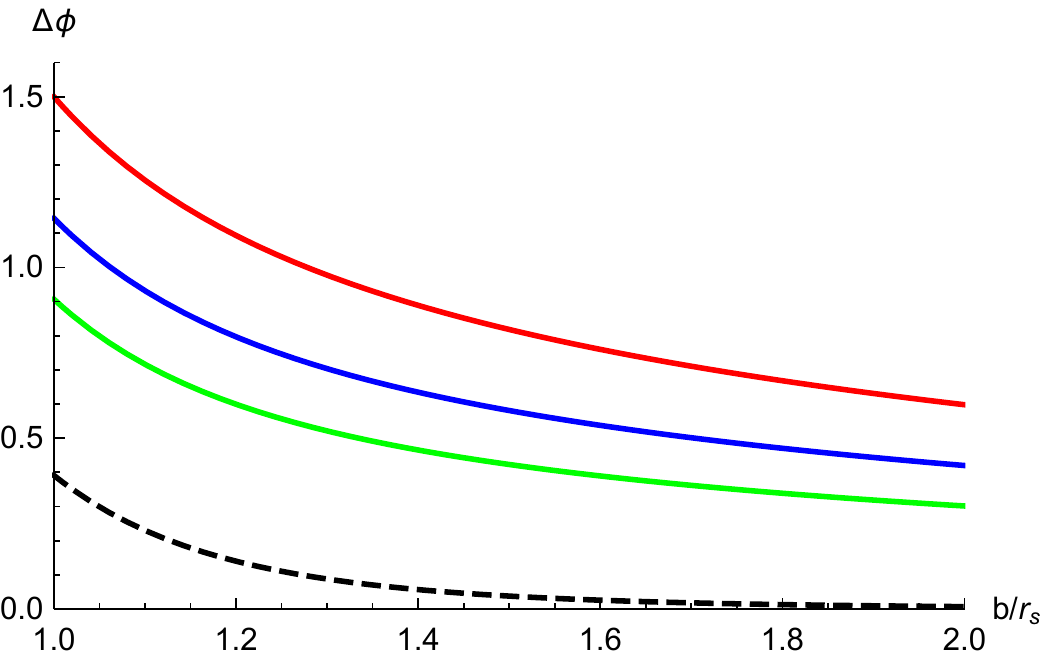}
\caption{The plots of $|\Delta \varphi_{\bot}|$ (dashed) and $\Delta \varphi$ (solid) for varying mass. 
Here $ M = 2 M_{\rm sun}, 1.4 M_{\rm sun}, M_{\rm sun}$ (top to bottom) and $\beta_0 = 5 \times 10^{11} \rm T$ with the same $r_s$ and $B_s$ as in Table 1. As $M \to 0$, the solid curve approaches to the dotted curve of Fig. 3, corresponding to the bending by Euler-Heisenberg Lagrangian.} \label{fig4}
\end{center}
\end{figure}

Here we consider the bending of light by magnetic dipole for the simple case when the light ray is passing on the equator. For a magnetic dipole located at the origin with arbitrary orientation, the bending can occur in both the horizontal and the vertical directions. Taking the direction of the incoming light as $x$-axis, the bending angle of each direction in the Euler-Heisenberg electrodynamics can be written in terms of the directional cosines ${\hat \alpha}= {\hat \mu} \cdot {\hat x} ,~ {\hat \beta}={\hat \mu} \cdot {\hat y},~ {\hat \gamma}= {\hat \mu} \cdot {\hat z}$ as \cite{KimLee2}
\be
 \Delta \varphi_{3, y} =  \frac{ \pi}{3 \cdot 2^7} a \frac{\alpha^2}{m_e^4} \left ( \frac{\hbar^3 \epsilon_0}{c^3} \right ) 
\left ( \frac{\mu_0}{4 \pi} \right )^2 \frac{\mu^2}{ b^6}  ( 15 {\hat \alpha}^2 + 41 {\hat \beta}^2 + 16 {\hat \gamma}^2 ),
\ee
\be
 \Delta \varphi_{3, z} =  \frac{5 \pi}{3 \cdot 2^6 } a \frac{\alpha^2}{m_e^4} \left ( \frac{\hbar^3 \epsilon_0}{c^3} \right ) 
\left ( \frac{\mu_0}{4 \pi} \right )^2 \frac{\mu^2}{ b^6} {\hat \beta}{\hat \gamma} ,
\ee
where $a =8$ ($14$) for the perpendicular (parallel) polazation. 

\section{Conclusion}

We consider the bending of light when it passes in the vicinity of a compact astrophysical object with strong magnetic field in the generalized Born-Infeld electrodynamics. To be specific, we compute the weak bending angle when the light is passing in the equatorial plane of the magnetic dipole. First we compute the bending angle using the trajectory equation based on geometric optics and show that it corresponds with the result from the Euler-Heisenberg Lagrangian in the appropriate limit. 

Then we consider the light bending by a massive object possessing magnetic dipole moment for the general relativistic computation. 
We compute the bending angle using the geodesic equation of the static axially symmetric solution for the metric and the four-potential. 
We obtain the bending angle as a function of impact parameter. In the massless limit,
 we confirm that the result agree with the bending angle obtained from the trajectory equation. 
As an application to astrophysics, we estimate the order of magnitude of the possible maximum bending angle for magnetars. 
When the impact parameter is large, the bending by mass term dominates. However, for magnetars with surface magnetic field of the order $10^{11} \rm T$,  the bending by nonlinear electromagnetic effects can be comparable to the bending by mass term  when the ray is passing close to the magnetar. There are other nonlinear electrodynamic consequences by such extremely intense field that can be comparable to those by mass. For example, it has been shown that the nonlinear electrodynamic shift from Euler-Heisenberg dipole for strong enough background field can reach values that are of the order of the gravitational redshift \cite{bonetti}. 

As mentioned before the classical Born-Infeld parameter $\beta$ is estimated much larger than the critical field of QED $B_c$. Then the nonlinear electrodynamic bending of light is essentially the same as the bending by Euler-Heisenberg Lagrangian. Because the bending angle obtained from the Euler-Heisenberg Lagrangian is polarization-dependent while the bending angle from general relativistic effect $\Delta \varphi_1$ is isotropic, birefringence of $\Delta \varphi_3$ may play an important role for observation. 

Recently astronomers, using the Event Horizon Telescope, have observed the signature of magnetic field at black holes \cite{akiyama}. If the magnetic field of black hole is intrinsic and strong enough, although this violates the no-hair theorem, one might consider the light bending by nonlinear electrodynamic effects. Close to the event horizon where the bending is strong one should compute the trajectory numerically from the geodesic equation. Not close to the event horizon where the bending is weak the same formalism used in this paper can be applied to compute the bending angle. We leave the extensions of our results to future work. 

\section*{Acknowledgements}
This work was supported by Basic Science Research Program through the National Research Foundation of Korea (NRF) funded by the Ministry of Education, Science and Technology (NRF-2019R1F1A1060409).

\end{document}